\documentclass[12pt, draftclsnofoot, onecolumn]{IEEEtran}

\usepackage{amssymb}
\usepackage{amsmath}
\usepackage{cite}
\usepackage{url}
\usepackage{cases}
\usepackage{empheq}
\usepackage{xcolor}
\usepackage{graphicx}
\usepackage{subfigure}
\usepackage{enumitem}
\usepackage{fancyhdr}
\usepackage{mdwmath}	
\usepackage{mdwtab}
\usepackage{caption}
\usepackage{amsthm}
\usepackage{algorithm}
\usepackage{algorithmic}

\newtheorem{lemma}{Lemma}
\newtheorem{remark}{Remark}
\newtheorem{theorem}{Theorem}
\newtheorem{corollary}{Corollary}
\newtheorem{assumption}{Assumption}
\newtheorem{definition}{Definition}

\newcommand{\eqr}[1]{(\ref{#1})}
\newcommand{\fref}[1]{Fig.~\ref{#1}}

\begin{document}
\title{Pinching-Antenna Systems for Physical Layer Security}
\author{Kaidi~Wang,~\IEEEmembership{Member,~IEEE,}
Zhiguo~Ding,~\IEEEmembership{Fellow,~IEEE,}
and Naofal~Al-Dhahir,~\IEEEmembership{Fellow,~IEEE}
\thanks{K. Wang and Z. Ding are with the Department of Electrical and Electronic Engineering, the University of Manchester, M1 9BB Manchester, U.K. (email: kaidi.wang@ieee.org; zhiguo.ding@ieee.org).}
\thanks{Z. Ding is also with Khalifa University, Abu Dhabi, UAE.}
\thanks{N. Al-Dhahir is with the Department of Electrical and Computer Engineering, University of Texas at Dallas, Richardson, TX 75080, USA (email: aldhahir@utdallas.edu).}}
\maketitle
\setlength{\abovedisplayskip}{2pt}
\setlength{\belowdisplayskip}{2pt}
\setlength{\textfloatsep}{0pt}
\begin{abstract}
This letter investigates the potential of pinching-antenna systems for enhancing physical layer security. By pre-installing multiple pinching antennas at discrete positions along a waveguide, the capability of the considered system to perform amplitude and phase adjustment is validated through the formulation of a secrecy rate maximization problem. Specifically, amplitude control is applied to enhance the signal quality at the legitimate user, while phase alignment is designed to degrade the received signal quality at the eavesdropper. This cooperation among pinching antennas is modeled as a coalitional game, and a corresponding antenna activation algorithm is proposed. The individual impact of each antenna is quantified based on the Shapley value and marginal contribution, providing a fair and efficient method for performance evaluation. Simulation results show that the considered pinching-antenna system achieves significant improvements in secrecy rate, and that the Shapley value based algorithm outperforms conventional coalition value based solutions.
\end{abstract}
\section{Introduction}
The ongoing evolution of sixth-generation (6G) wireless communications is driving the development of advanced antenna technologies capable of supporting higher data rates, improved reliability, and adaptability to dynamic environments. In this context, flexible antennas have emerged as a promising solution, offering reconfigurability beyond the constraints of conventional fixed-location designs. Among the flexible-antenna technologies, fluid antennas and movable antennas have received considerable attention, with the former altering positions or shapes through reconfigurable liquid structures, and the latter physically shifting within a defined boundary \cite{wong2020fluid, zhu2023modeling}. While these approaches provide valuable flexibility, inherent limitations in wavelength-scale adjustments reduce the effectiveness in complex and dynamic wireless environments. To overcome these limitations, pinching antennas have been proposed recently as a novel type of flexible antennas \cite{ding2024pin}. By applying small dielectric particles along a dielectric waveguide, the pinching-antenna system is able to dynamically construct line-of-sight (LoS) links and/or enhance transceiver channels, thereby mitigating large-scale path loss and supporting adaptable wireless channels \cite{kaidi2025pin, ding2025blockage}.

As a promising research direction in pinching-antenna systems, physical layer security has been extensively explored in recent works \cite{badarneh2025physical, jiang2025pinching, papanikolaou2025secrecy, zhu2025pinching, sun2025physical}. In \cite{badarneh2025physical}, the physical layer security performance of a pinching-antenna system was analyzed in a single-antenna scenario, with various secrecy performance metrics derived and validated through simulations. To maximize the data rate of the legitimate user while constraining the error rate of the eavesdropper, both single-waveguide and multi-waveguide pinching-antenna systems were examined under position and noise uncertainties \cite{jiang2025pinching}. The study in \cite{papanikolaou2025secrecy} incorporated artificial noise to enhance security, addressing secrecy rate maximization in a single-antenna scenario through distinct approaches for single-waveguide and multi-waveguide configurations. In \cite{zhu2025pinching}, the secrecy rate maximization problem was extended to a multi-antenna scenario by integrating artificial noise and beamforming design for both single- and multi-waveguide systems. Similarly, \cite{sun2025physical} considered the multi-antenna case, where iterative optimization of antenna placement and beamforming was employed to improve secrecy performance in both single-user and multi-user scenarios.

In the aforementioned studies, a variety of alternative techniques have been employed to enhance the physical layer security of pinching-antenna systems, such as artificial noise generation, beamforming design, and power control. Despite these efforts, the inherent potential of pinching antennas in this context has not yet been fully exploited, which motivates this letter. In this work, by considering pre-installed pinching antennas, a secrecy rate maximization problem is investigated, revealing that the collaboration among pinching antennas enables joint adjustment of both amplitude and phase, thereby improving the achievable secrecy rate. To effectively characterize this cooperation, a coalitional game based antenna activation algorithm is proposed, in which the impact of each pinching antenna is quantified based on the Shapley value and marginal contribution. Simulation results validate the effectiveness of the proposed strategy, demonstrating that the pinching-antenna system outperforms conventional fixed-location antenna systems, with the proposed algorithm achieving significant improvements in secrecy rate performance.
\section{System Model and Problem Formulation}
\begin{figure}[!t]
\centering{\includegraphics[width=90mm]{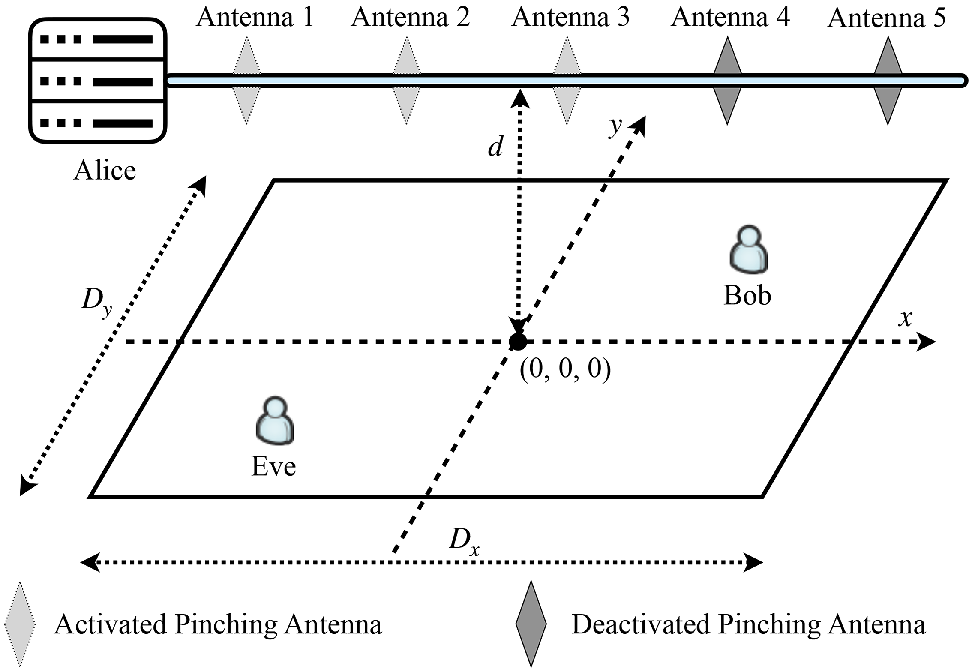}}
\caption{An illustration of the considered pinching-antenna system in physical layer security scenarios with $N=5$ antenna positions. The set of activated antennas is $\mathcal{S}=\{4, 5\}$.}
\label{system}
\end{figure}

Consider a downlink pinching-antenna system with one base station (Alice), one user (Bob) and one eavesdropper (Eve), where Alice transmits the signal to Bob via $N$ preconfigured pinching antennas on the waveguide. As shown in \fref{system}, this system is within a rectangular region with dimensions $D_x$ and $D_y$, and the waveguide is deployed at a height of $d$. The collection of all pinching antennas is $\mathcal{N}=\{1,2,\cdots, N\}$. The locations of pinching antenna $n$, Bob, and Eve are denoted by $\boldsymbol{\psi}_n^\mathrm{Pin}=(x_n^\mathrm{Pin},0,d)$, $\boldsymbol{\psi}^\mathrm{Bob}=(x^\mathrm{Bob},y^\mathrm{Bob},0)$, and $\boldsymbol{\psi}^\mathrm{Eve}=(x^\mathrm{Eve},y^\mathrm{Eve},0)$, respectively. Moreover, it is assumed that the locations of Bob and Eve are available at Alice.
\subsection{Signal Model}
In the considered pinching-antenna system, by including the phase shifts caused by signal propagation within the dielectric waveguide \cite{ding2024pin}, the complex channel coefficient between antenna $n$ and Bob can be expressed as follows:
\begin{equation}
h_n^\mathrm{Bob} = \frac{\eta e^{-j\left(\frac{2\pi}{\lambda}\left\|\boldsymbol{\psi}^\mathrm{Bob}-\boldsymbol{\psi}_n^\mathrm{Pin}\right\|+\frac{2\pi}{\lambda_g}\left\|\boldsymbol{\psi}_0^\mathrm{Pin}\!-\boldsymbol{\psi}_n^\mathrm{Pin}\right\|\right)}}{\left\|\boldsymbol{\psi}^\mathrm{Bob}-\boldsymbol{\psi}_n^\mathrm{Pin}\right\|},
\end{equation}
where $\eta=\frac{c}{4\pi f_c}$ is the free-space path loss, $c$ is the speed of light, $f_c$ is the carrier frequency, $\lambda$ is the wavelength, $\lambda_g=\lambda/n_\mathrm{eff}$ is the wavelength in the waveguide, $n_\mathrm{eff}$ is the effective refractive index of the dielectric waveguide \cite{pozar2021microwave}, $\|\boldsymbol{\psi}^\mathrm{Bob}-\boldsymbol{\psi}_n^\mathrm{Pin}\|$ is the distance between Bob and pinching antenna $n$, and $\|\boldsymbol{\psi}_0^\mathrm{Pin}\!-\boldsymbol{\psi}_n^\mathrm{Pin}\|$ is the distance between the feed point and pinching antenna $n$.

Based on the antenna activation indicator $\alpha_n\in\{0,1\}$, the channel vector between Alice and Bob through all activated pinching antenna can be expressed as follows:
\begin{equation}
\mathbf{h}^\mathrm{Bob}=\left[\alpha_1h_1^\mathrm{Bob} \quad \alpha_2h_2^\mathrm{Bob} \quad\cdots\quad \alpha_N h_N^\mathrm{Bob}\right]^\top,
\end{equation}
where $(\cdot)^\top$ denotes the transpose operation. In the above equation, $\alpha_n=1$ indicates that pinching antenna $n$ is activated, $\alpha_n=0$ otherwise. Similarly, the channel vector between Alice and Eve is given by
\begin{equation}
\mathbf{h}^\mathrm{Eve}=\left[\alpha_1h_1^\mathrm{Eve} \quad \alpha_2h_2^\mathrm{Eve} \quad\cdots\quad \alpha_N h_N^\mathrm{Eve}\right]^\top,
\end{equation}
where $h_n^\mathrm{Eve}$ is the complex channel coefficient between antenna $n$ and Eve, i.e., 
\begin{equation}
h_n^\mathrm{Eve} = \frac{\eta e^{-j\left(\frac{2\pi}{\lambda}\left\|\boldsymbol{\psi}^\mathrm{Eve}-\boldsymbol{\psi}_n^\mathrm{Pin}\right\|+\frac{2\pi}{\lambda_g}\left\|\boldsymbol{\psi}_0^\mathrm{Pin}\!-\boldsymbol{\psi}_n^\mathrm{Pin}\right\|\right)}}{\left\|\boldsymbol{\psi}^\mathrm{Eve}-\boldsymbol{\psi}_n^\mathrm{Pin}\right\|}.
\end{equation}

Assuming that the transmit power is equally allocated to all antennas and the waveguide propagation loss is omitted \cite{kaidi2025pin2}, the data rates at Bob and Eve can presented as follows:
\begin{equation}\label{rate}
R^\mathrm{x}=\log_2\!\!\left(1\!+\!\rho\left|\sum_{n=1}^N \alpha_n h_n^\mathrm{x}\right|^2\right), \forall \mathrm{x}\in\{\mathrm{Bob}, \mathrm{Eve}\},
\end{equation}
where $\rho=P_t/(K\sigma^2)$ is the transmit signal-to-noise ratio (SNR), $P_t$ is the transmit power, $K=\sum_{n=1}^N\alpha_n$ is the number of activated pinching antennas, and $\sigma^2$ is the noise power. As a result, the secrecy data rate is given by
\begin{equation}
R =R^\mathrm{Bob}-R^\mathrm{Eve}.
\end{equation}

\subsection{Problem Formulation}
Based on antenna activation, a secrecy rate maximization problem is formulated in order to improve the considered pinching-antenna system in the physical layer security scenario, as follows:
\begin{subequations}
\begin{empheq}{align}
\max_{\boldsymbol{\alpha}}\quad & R\\
\textrm{s.t.} \quad & \alpha_n \in\{0,1\}, \forall n\in\mathcal{N},\\
& \sum\nolimits_{n=1}^N\alpha_n\ge 1,
\end{empheq}
\label{problem}
\end{subequations}\vspace{-2mm}\\
where $\boldsymbol{\alpha}$ is the collection of all antenna activation indicators. Constraint (\ref{problem}c) indicates that at least one pinching antenna is activated on the waveguide.
\section{Coalitional Game based Antenna Activation}
In this section, the effectiveness of pinching-antenna systems in improving physical layer security is investigated, and a coalitional game based algorithm is proposed to solve the formulated antenna activation problem.
\subsection{Feasibility Analysis}

\begin{figure}[!t]
\centering{\includegraphics[width=90mm]{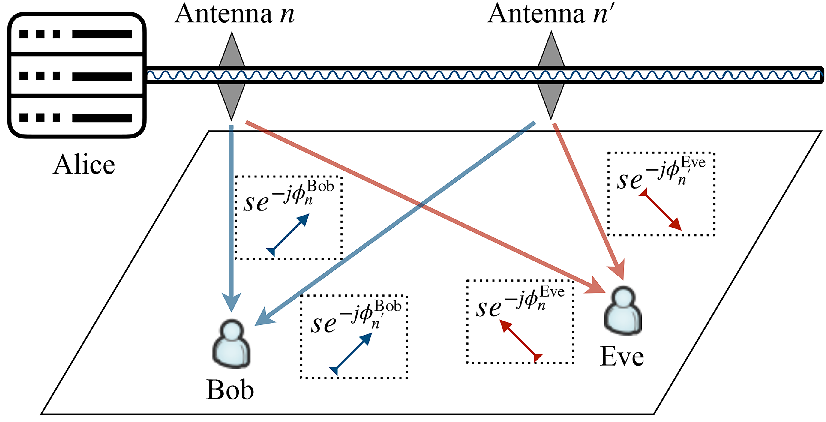}}
\caption{An illustration of phase adjustment via two-antenna activation in the considered pinching-antenna system.}
\label{example}
\end{figure}

It is evident that problem \eqr{problem} can be addressed by jointly considering two complementary strategies, including enhancing the data rate at Bob and suppressing the data rate at Eve. As indicated in \eqr{rate}, the achievable rates for both Bob and Eve are dictated by their respective effective channels, which result from the coherent summation of multiple complex-valued channel coefficients. In this context, the precise phase adjustment of the individual channels can be achieved through the activation of multiple pinching antennas \cite{zhu2025pinching}. Such phase control facilitates the constructive or destructive combination of signals, thereby enhancing or attenuating the effective channels in accordance with specific performance objectives.

Consider a two-antenna scenario as an illustrative example. To maximize the channel amplitude, the first pinching antenna, denoted by $n$, is activated at the position closest to Bob, as shown in \fref{example}. The second pinching antenna, denoted by $n'$, is activated to enable phase control. In this configuration, the effective channel at Bob can be maximized if
\begin{equation}\label{con1}
\text{mod}\left\{\phi_n^\mathrm{Bob}-\phi_{n'}^\mathrm{Bob},2\pi\right\}=0,
\end{equation}
where $\text{mod}\{a, b\}$ denotes the modulo operation of $a$ by $b$, and
\begin{equation}
\phi_n^\mathrm{x}\triangleq\!\frac{2\pi}{\lambda}\left\|\boldsymbol{\psi}^\mathrm{x}\!-\!\boldsymbol{\psi}_n^\mathrm{Pin}\right\|+\frac{2\pi}{\lambda_g}\left\|\boldsymbol{\psi}_0^\mathrm{Pin}\!-\!\boldsymbol{\psi}_n^\mathrm{Pin}\right\|, \forall \mathrm{x}\!\in\!\{\mathrm{Bob}, \mathrm{Eve}\}.
\end{equation}
Meanwhile, if pinching antenna $n'$ produces a phase difference of $\pi$ relative to the signal transmitted through antenna $n$ at Eve, i.e.,
\begin{equation}\label{con2}
\text{mod}\left\{\phi_n^\mathrm{Eve}-\phi_{n'}^\mathrm{Eve},2\pi\right\}=\pi,
\end{equation}
the effective channel at Eve will be significantly attenuated. In this example, the objective is to identify antenna positions that simultaneously satisfy conditions \eqr{con1} and \eqr{con2}.

\begin{remark}
In the considered system, the activation of subsequent pinching antennas beyond the initial one may either enhance or degrade the data rate, depending on phase differences that determine constructive or destructive interference.
\end{remark}

\begin{remark}
Due to the spatial separation of users, antenna activation can selectively influence effective channels based solely on position information, thereby enabling a practical approach to enhancing physical layer security.
\end{remark}

\begin{remark}
The current pinching-antenna framework can be extended to multi-waveguide scenarios, where phase control facilitates inter-waveguide interference management. This additional flexibility enables interference mitigation and signal optimization, potentially improving system capacity.
\end{remark}

By identifying the potential of phase based effective channel adjustment, additional pinching antennas can be activated to maximize the secrecy data rate. However, as the number of antennas increases, the complexity of determining the optimal antenna configuration rises significantly. Moreover, improving system performance requires the joint consideration of both amplitude and phase, which further amplifies the complexity. In this work, this challenge is addressed by selecting an appropriate subset from the pre-installed pinching antennas.
\subsection{Design of Coalitional Game based Algorithm}
As previously discussed, the core objective in solving problem \eqr{problem} is to identify the optimal combination of activated pinching antennas. This task can be formulated as a coalitional game with non-transferable utility (NTU) \cite{han2012game}, denoted by $(\mathcal{N}, v)$, where the activated pinching antennas form a coalition $\mathcal{S}$ with the aim of maximizing the coalition value $v(\mathcal{S})$, i.e., the secrecy data rate. Subject to the constraints of problem \eqr{problem}, the coalition satisfies $\mathcal{S}\subseteq\mathcal{N}$ and $|\mathcal{S}|\ge 1$.

In the considered system, the secrecy data rate is achieved through the collaboration of all activated pinching antennas. Specifically, each antenna contributes individually by enhancing the channel amplitude while simultaneously participating in phase based coordination with others. Due to this contribution mechanism, fair allocation of individual payoffs becomes challenging. Accordingly, the Shapley value is adopted as a solution concept to quantify the impact of each pinching antenna within the coalition. Given the coalition value $v(\mathcal{S})$, the payoff of pinching antenna $n\in\mathcal{S}$ is defined as
\begin{equation}
\varphi_n(\mathcal{S},v)=\!\!\!\!\!\sum_{\mathcal{S'}\subseteq\mathcal{S}\backslash\{n\}}\!\!\!\!\!\frac{|\mathcal{S'}|!(|\mathcal{S}|\!-\!|\mathcal{S}'|\!-\!1)!}{|\mathcal{S}|!}\left[v(\mathcal{S}'\cup\{n\})\!-\!v(\mathcal{S}')\right].
\end{equation}

In the case that pinching antenna $n$ is not a member of coalition $\mathcal{S}$, the Shapley value cannot be applied. However, this antenna may still influence the coalition. For instance, if an antenna exerts a negative impact when included in coalition $\mathcal{S}$, this effect can be interpreted as its potential contribution when excluded from the coalition. By evaluating its impact on coalition $\mathcal{S}$, the hypothetical payoff of any antenna $n\notin\mathcal{S}$ can be defined as its marginal contribution, as follows:
\begin{subequations}
\label{outside}
\begin{empheq}[left={\varphi_n(\mathcal{N}\backslash\mathcal{K},v)=}\empheqlbrace]{align}
& v(\mathcal{S})\!-\!v(\mathcal{S}\cup\{n\}), &\!\!\!\!\!\text{if} \quad n\notin\mathcal{S},\\
& v(\mathcal{S}\backslash \{n\})\!-\!v(\mathcal{S}), &\!\!\!\!\!\text{if} \quad n\in\mathcal{S}.
\end{empheq}
\end{subequations}
In the above equation, (\ref{outside}a) and (\ref{outside}b) correspond to marginal contributions of antenna $n$ upon joining and leaving coalition $\mathcal{S}$, respectively.

Based on the payoff definition, the strategy of each pinching antenna with respect to a given coalition $\mathcal{S}$ can be determined. Specifically, any pinching antenna may decide to join or leave coalition $\mathcal{S}$ to maximize its payoff. This process follows the merge-and-split strategy \cite{kaidi2024nfc}, as defined below:
\begin{definition}
\textbf{(Merge Rule)} Given a coalition $\mathcal{S}$, any pinching antenna $n\notin\mathcal{S}$ tends to merge into $\mathcal{S}$ if and only if $\varphi_n(\mathcal{S},v) >\varphi_n(\mathcal{N}\backslash\mathcal{K},v)$.
\end{definition}

\begin{definition}
\textbf{(Split Rule)} Given a coalition $\mathcal{S}$, any pinching antenna $n\in\mathcal{S}$ tends to split from $\mathcal{S}$ if and only if $\varphi_n(\mathcal{N}\backslash\mathcal{K},v) > \varphi_n(\mathcal{S},v)$ and $|\mathcal{S}|\neq 1$.
\end{definition}

\begin{algorithm}[t]
\caption{Coalitional Game based Antenna Activation}
\label{alg}
\begin{algorithmic}[1]
\STATE \textbf{Initialization:}
\STATE Activate the closet pinching antenna to Bob to obtain $\mathcal{S}$.
\STATE \textbf{Main Loop:}
\FOR{$n\in\mathcal{N}$}
\IF{$N\notin \mathcal{S}$ and Merge Rule holds}
\STATE $\mathcal{S}=\mathcal{S}\cup\{n\}$.
\ENDIF
\IF{$N\in \mathcal{S}$ and Split Rule holds}
\STATE $\mathcal{S}=\mathcal{S}\backslash\{n\}$.
\ENDIF 
\ENDFOR
\end{algorithmic}
\end{algorithm}

Based on the definitions of merge and split rules, a coalitional game based antenna activation algorithm is proposed in Algorithm~\ref{alg}. This algorithm terminates when no pinching antenna can be activated or deactivated during a complete cycle, i.e., within lines 4-11. At this point, no pinching antenna has an incentive to deviate from the current coalition structure, thereby confirming Nash stability, as defined below:
\begin{definition}\label{stable}
In coalitional game $(\mathcal{N},v)$, a coalition $\mathcal{S}$ is Nash stable if and only if
\begin{enumerate}
\item $\varphi_n(\mathcal{S},v) \le \varphi_n(\mathcal{N}\backslash\mathcal{K},v), \forall n \notin\mathcal{S}$; and
\item $\varphi_n(\mathcal{N}\backslash\mathcal{K},v) \le \varphi_n(\mathcal{S},v), \forall n \in\mathcal{S}$.
\end{enumerate}
\end{definition}
\subsection{Properties of Coalitional Game based Algorithm}
In Algorithm~\ref{alg}, all pinching antennas must assess whether to merge into or split from coalition $\mathcal{S}$ in each cycle. Given a total of $C$ cycles, the computational complexity of the proposed algorithm is $\mathcal{O}(CN)$, where $\mathcal{O}$ is the big-O notation. Furthermore, computing the Shapley value leads to a complexity of $\mathcal{O}(|\mathcal{S}|\times 2^{|\mathcal{S}|})$, as it involves evaluating the contribution of each antenna in coalition $\mathcal{S}$ across $2^{|\mathcal{S}|-1}$ possible subsets. In the considered system, the coalition formation is constrained by specific phase alignment requirements, which inherently limits the size of coalition $\mathcal{S}$ and thereby reduces the overall computational cost.

Since the payoff of each pinching antenna is strictly increasing during the execution of Algorithm~\ref{alg}, the proposed algorithm is guaranteed to converge to a final coalition structure within a finite number of iterations. In this final structure, no pinching antenna can improve its payoff by unilaterally deviating, as defined in Definition~\ref{stable}. Therefore, the resulting coalition structure is always Nash stable.

In the considered pinching-antenna system, due to the underlying phase adjustment mechanism, the proposed payoff based solution can outperform conventional methods that rely on coalition value. For example, consider a coalition $\mathcal{S}=\{n, n'\}$ in which perfect phase alignment is achieved, such that conditions \eqref{con1} and \eqref{con2} are satisfied. In this scenario, the activation of additional antennas would disrupt the established phase balance, leading to their exclusion from $\mathcal{S}$. However, the Shapley value based approach enables the evaluation of individual antenna contributions, thereby allowing the potential inclusion of additional pinching antennas. Through this process, perfect phase alignment can be re-established either by deactivating antennas with lower payoffs or by activating additional antennas. Consequently, a more optimal coalition structure can be constructed.

\section{Simulation Results}
In this section, the performance of the considered pinching-antenna system and proposed solution is presented. In this simulation, Bob and Eve are randomly distributed within a defined rectangular region, and the waveguide is deployed along the line $y_n^\mathrm{Pin}=0$ with a length of $10$~meters and a height of $d=3$~meters. The pre-installed pinching antennas are uniformly placed along the waveguide. The simulation parameters are configured as follows: $D_x=10$~meters, $D_y=6$~meters, $f_c=28$~GHz, $n_\mathrm{eff}=1.4$, and $\sigma^2=-90$~dBm. Furthermore, a fixed-location antenna system is employed as the benchmark, where a uniform linear array consisting of $N$ antennas with the antenna space of $\frac{\lambda}{2}$ is positioned at the center of the region. 

\begin{figure}[!t]
\centering{
\subfigure{\centering{\includegraphics[width=80mm]{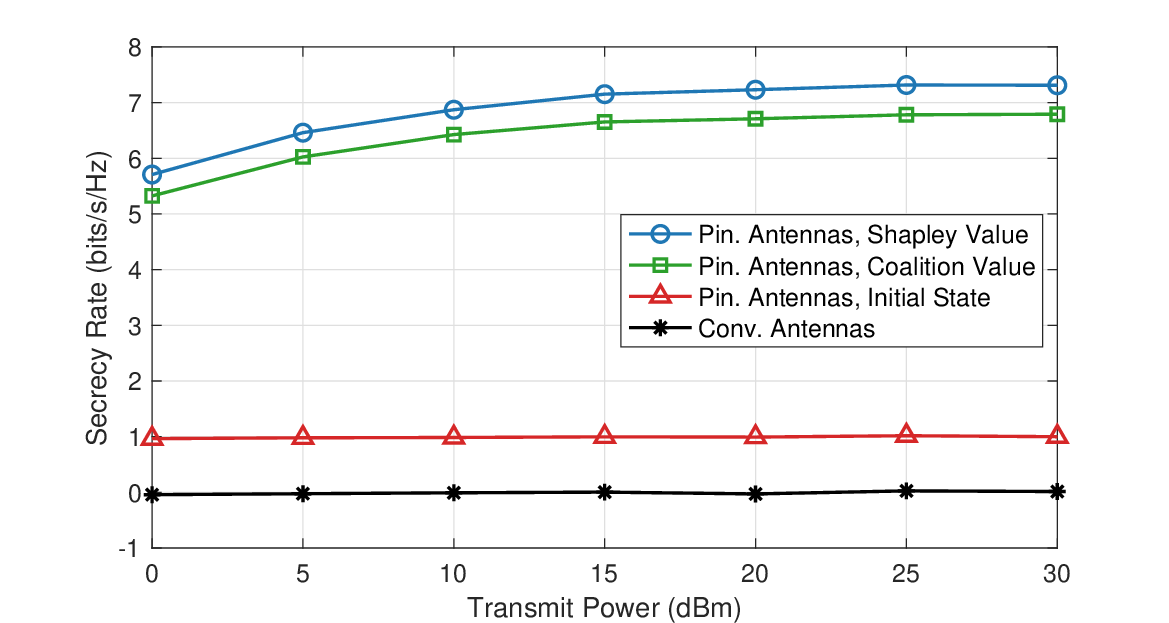}}}
\subfigure{\centering{\includegraphics[width=80mm]{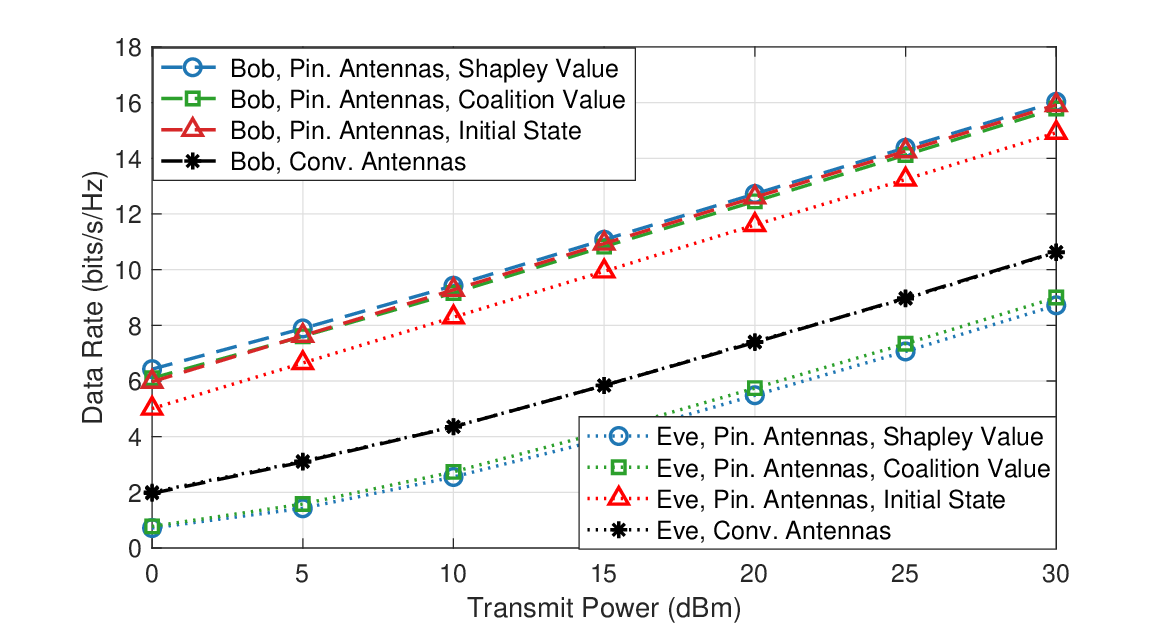}}}}
\caption{Impact of the transmit power on the secrecy rate and the individual data rate, where $N=20$.}
\label{result1}
\end{figure}

In \fref{result1}, the effect of transmit power on the secrecy rate and the corresponding individual data rates is presented. It can be observed that the proposed coalitional game based algorithm can substantially enhance the secrecy rate, and this improvement is increasing with transmit power. In contrast, the secrecy rate of the conventional fixed-location antenna system is close to zero, as the individual data rates at both Bob and Eve are nearly identical. For the initial state, a single pinching antenna is activated at the position closest to Bob, resulting in a slightly higher data rate at Bob compared to Eve and achieving a secrecy rate of approximately1 bit/s/Hz. By applying the proposed coalitional game based algorithm, the data rate at Bob is marginally improved, while the data rate at Eve is significantly suppressed, thereby improving the secrecy rate. Moreover, based on the Shapley value, the proposed algorithm can be performed more efficiently, leading to a further enhancement in the secrecy rate.

\begin{figure}[!t]
\centering{
\subfigure{\centering{\includegraphics[width=80mm]{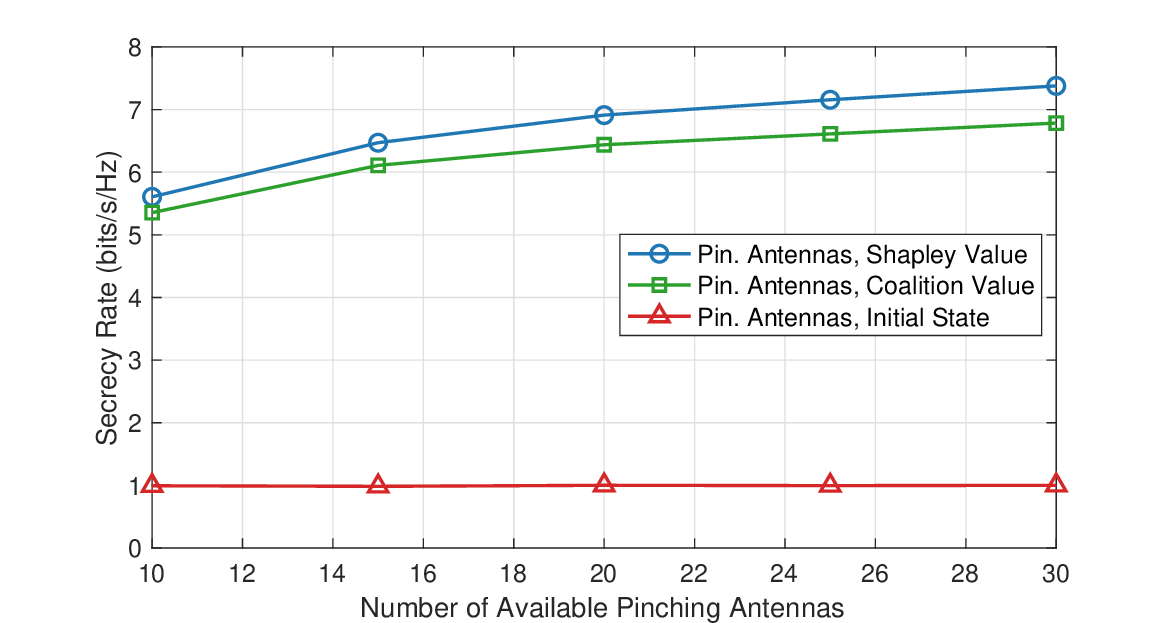}}}
\subfigure{\centering{\includegraphics[width=80mm]{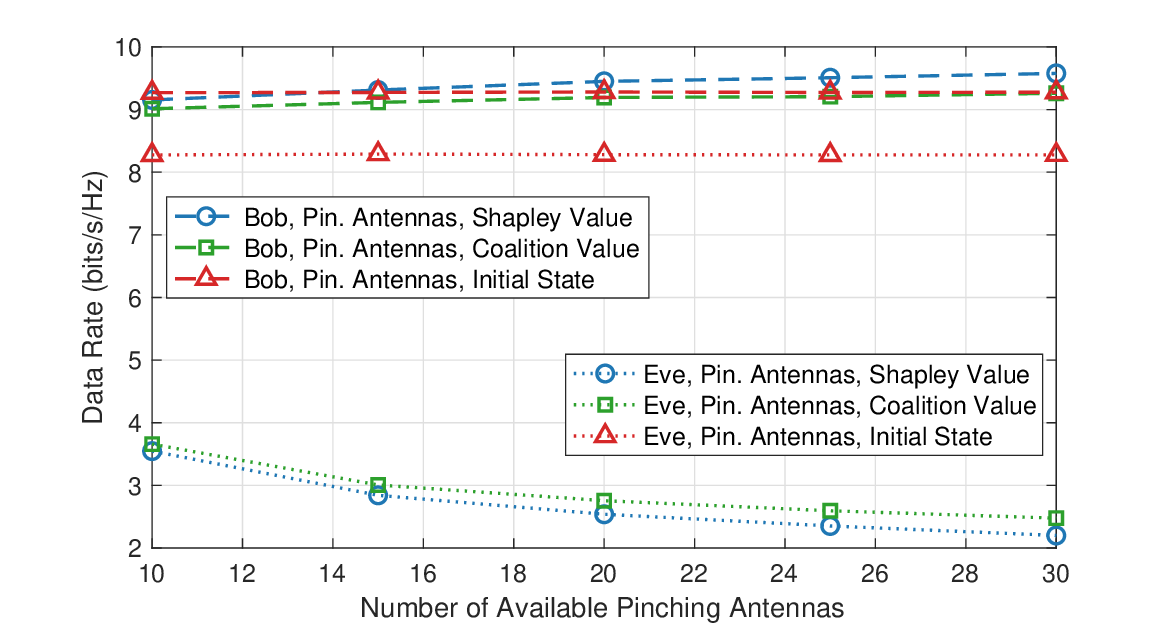}}}}
\caption{Impact of the number of antennas on the secrecy rate and the individual data rate, where $P_t=10$~dBm.}
\label{result2}
\end{figure}

In \fref{result2}, the number of available pinching antennas is increased to demonstrate its impact. With more available pinching antennas, antenna activation can be adjusted more precisely, resulting in enhanced secrecy performance. As shown in \fref{result2}, the secrecy rate improves with the number of antennas, which is due to an increase in Bob's achievable data rate and a decrease in Eve's data rate. Specifically, the improvement in Bob's data rate is relatively limited, whereas the degradation in Eve's data rate is more significant. This result reflects the different adjustment strategies, with amplitude control playing a key role for Bob and phase alignment being the primary factor affecting Eve. The availability of additional pinching antennas enables more accurate phase alignment, thereby strengthening the intended interference at Eve. Furthermore, it is worth noting that the Shapley value based algorithm can achieve a higher secrecy rate, characterized by an increased data rate at Bob and a decreased data rate at Eve. In the initial state, since only one pinching antenna is activated, its performance is not affected, including the secrecy rate and individual data rates. 

\begin{figure}[!t]
\centering{
\subfigure{\centering{\includegraphics[width=80mm]{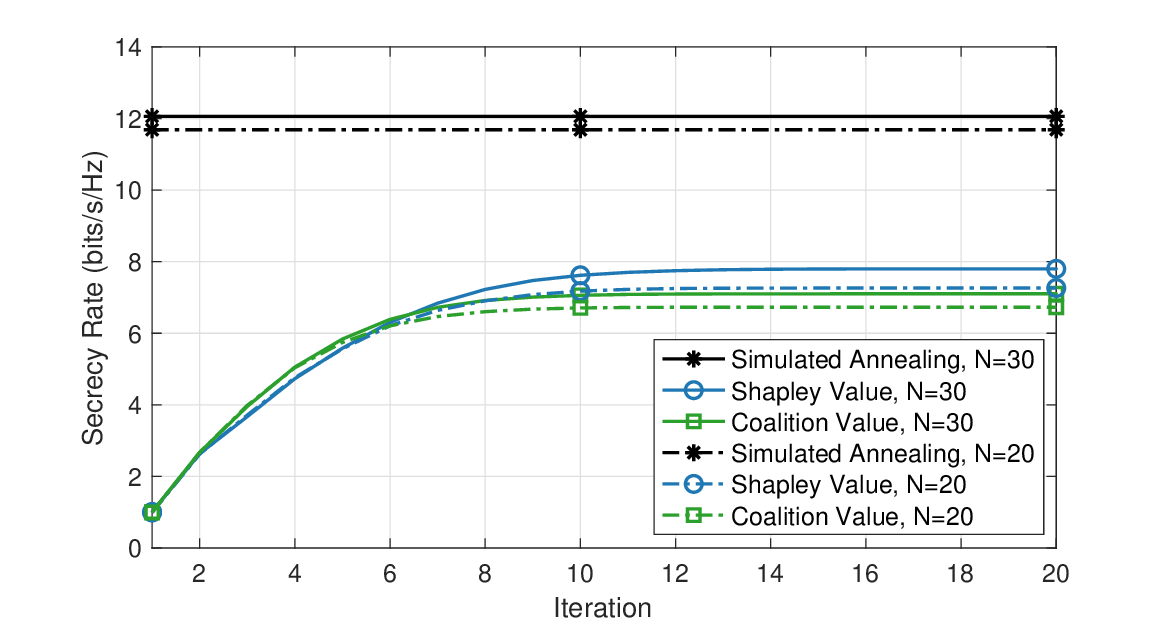}}}
\subfigure{\centering{\includegraphics[width=80mm]{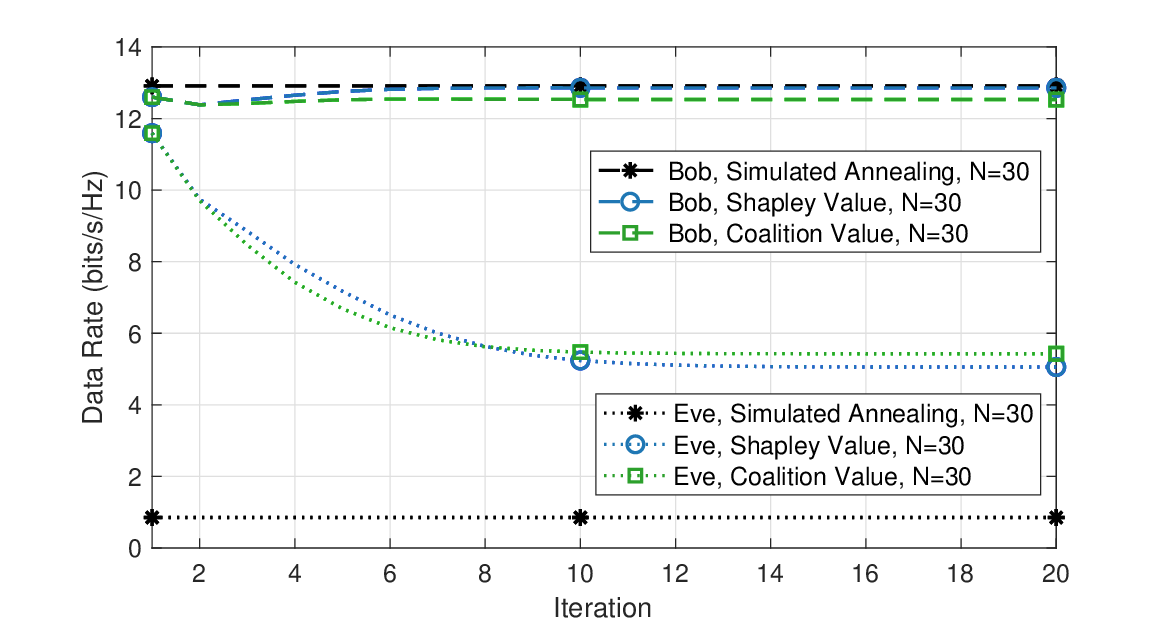}}}}
\caption{Convergence performance of the proposed algorithm, where $P_t=20$~dBm.}
\label{result3}
\end{figure}

The convergence performance of the proposed coalitional game based algorithm is illustrated in \fref{result3}, where a simulated annealing method is utilized to compute the global optimal solution. The results indicate that, in comparison with the conventional coalition value based algorithm, the proposed Shapley value based algorithm can converge to a superior coalition structure, although it requires more iterations. Due to the inherent complexity of achieving perfect phase alignment, the proposed algorithm attains approximately $65\%$ of the global optimum. However, in contrast to the simulated annealing method, which requires a total of $10^6$ iterations to reach the global optimum, the proposed coalitional game based algorithm achieves comparable performance with only $15$ iterations. Furthermore, it can be observed that the primary source of the performance gap is the data rate at Eve, which can be suppressed to below $1$~bit/s/Hz under the global optimal solution. This observation reinforces the conclusion that attaining perfect phase alignment remains a considerable practical challenge.

\section{Conclusions}
In this work, the capability of pinching-antenna systems in terms of physical layer security was explored through the formulation of a secrecy rate maximization problem. By activating the pre-installed pinching antennas, joint amplitude and phase adjustments were employed to simultaneously improve the data rate for the legitimate user while degrading it for the eavesdropper. To comprehensively evaluate the individual impact of each pinching antenna, the concepts of Shapley value and marginal contribution were integrated into a proposed coalitional game based algorithm. Simulation results confirmed the effectiveness of the phase alignment strategy and demonstrated the performance gains achieved by the proposed solution.
\bibliographystyle{IEEEtran}
\bibliography{KaidisBib}
\end{document}